\documentclass[amssymb,prd,superscriptaddress,aps,nofootinbib,twocolumn,showpacs,10pt]{revtex4-1}
\usepackage{graphicx, epsfig, amssymb} 
\usepackage{amsmath, amsfonts}
\usepackage{bm} 
\usepackage[breaklinks]{hyperref}
\usepackage{color}
\usepackage{enumerate}

\def\nn{\nonumber}

\def\be{\begin{equation}}
\def\ee{\end{equation}}
\def\beq{\begin{eqnarray}}
\def\eeq{\end{eqnarray}}

\begin{document}

\title{Exotic Compact Objects and How to Quench their Ergoregion Instability}

\author{Elisa Maggio}\email{elisa.maggio@roma1.infn.it}
\affiliation{Dipartimento di Fisica, ``Sapienza'' Universit\`a di Roma, Piazzale Aldo Moro 5, 00185, Roma, Italy.}
\affiliation{Sezione INFN Roma1, Piazzale Aldo Moro 5, 00185, Roma, Italy.}
\author{Paolo Pani}\email{paolo.pani@roma1.infn.it}
\affiliation{Dipartimento di Fisica, ``Sapienza'' Universit\`a di Roma, Piazzale Aldo Moro 5, 00185, Roma, Italy.}
\affiliation{Sezione INFN Roma1, Piazzale Aldo Moro 5, 00185, Roma, Italy.}
\affiliation{CENTRA, Departamento de F\'{\i}sica, Instituto Superior T\'ecnico, Universidade de Lisboa, Avenida~Rovisco Pais 1, 1049 Lisboa, Portugal.}
\author{Valeria Ferrari}\email{valeria.ferrari@roma1.infn.it}
\affiliation{Dipartimento di Fisica, ``Sapienza'' Universit\`a di Roma, Piazzale Aldo Moro 5, 00185, Roma, Italy.}
\affiliation{Sezione INFN Roma1, Piazzale Aldo Moro 5, 00185, Roma, Italy.}

\begin{abstract}
Gravitational-wave astronomy can give us access to the structure of black holes, potentially probing microscopic or even Planckian corrections at the horizon scale, as those predicted by some quantum-gravity models of exotic compact objects. A generic feature of these models is the replacement of the horizon by a reflective surface. Objects with these properties are prone to the so-called ergoregion instability when they spin sufficiently fast.
We investigate in detail a simple model consisting of scalar perturbations of a Kerr geometry with a reflective surface near the horizon. The instability depends on the spin, on the compactness, and on the reflectivity at the surface.
The instability time scale increases only logarithmically in the black-hole limit and, for a perfectly reflecting object, this is not enough to prevent the instability from occurring on dynamical time scales.
However, we find that an absorption rate at the surface as small as $0.4\%$ (reflectivity coefficient as large as $|{\cal R}|^2=0.996$) is sufficient to quench the instability completely.
Our results suggest that exotic compact objects are not necessarily ruled out by the ergoregion instability.
\end{abstract}

\maketitle

\section{Introduction}

Several arguments have recently been put forward, suggesting that new physics at the horizon scale might alter or even halt the formation of black holes (BHs) during the gravitational collapse~\cite{Mazur:2001fv,Mathur:2005zp,Skenderis:2008qn,Almheiri:2012rt,Giddings:2014ova}.
In this context, several models of exotic compact objects (ECOs) have been proposed as alternatives to BHs~\cite{Holdom:2016nek,Brustein:2017kcj,Barcelo:2017lnx} or simply as exotic gravitational sources. In both cases, ECOs will be smoking guns for new physics at the poorly-explored horizon scale. While different models predict objects with rather different properties, all ECOs have some features in common: their mass and compactness can be arbitrarily close to those of a BH, and they do not possess an event horizon.

The horizon scale is extremely challenging --~if not impossible~\cite{Abramowicz:2002vt}~-- to probe through electromagnetic observations, but gravitational-wave (GW) astronomy gives us access to the very structure of BHs, carrying unique information on the dynamical processes that lead to the formation of an event horizon.

Following the recent detection of GWs from compact-binary coalescences~\cite{Abbott:2016blz,Abbott:2016nmj}, there has been a growing interest in detecting or ruling out ECOs through GW observations. It has been recently
argued that putative corrections at the horizon scale will appear as GW echoes in the postmerger ringdown phase of a binary coalescence~\cite{Cardoso:2016rao,Cardoso:2016oxy} (see also Ref.~\cite{Ferrari:2000sr} for an earlier study, and Refs.~\cite{Abedi:2016hgu,Ashton:2016xff,Abedi:2017isz} for a debate on the evidence of this effect in the aLIGO data). Likewise, the measurement of the tidal deformability of the two objects~\cite{Wade:2013hoa,Cardoso:2017cfl}, or of their spin-induced quadrupole moment~\cite{Krishnendu:2017shb}, during the late-time inspiral of the coalescence can also be used to distinguish ECOs from BHs (for related work on other GW signatures of ECOs, cf.\ Refs.~\cite{Pani:2009ss,Macedo:2013jja,Giudice:2016zpa,Chirenti:2016hzd}).

Most of these studies have been focused on static objects, neglecting the spin or including it in a simplistic way. Nonetheless, the role of the spin is crucial for ECOs, because highly-spinning, compact objects might turn unstable due to the so-called \emph{ergoregion instability}~\cite{1978CMaPh..63..243F} (for a review, see Ref.~\cite{Superradiance}). The latter is an instability that develops in any spacetime featuring a finite\footnote{A counterexample to the instability is provided when the ergoregion extends all the way to infinity as in certain nonasymptotically flat geometries~\cite{Dias:2009ex,Dias:2012pp}.} ergoregion but without an event horizon: since physical negative-energy states can exist inside the ergoregion --~which is key ingredient of the Penrose's process~\cite{Penrose:1969}~-- it is energetically favorable to cascade toward even more negative states. The only way to prevent such infinite cascade from developing is by absorbing the negative-energy states. Kerr BHs can absorb radiation very efficiently and are indeed stable even if they have an ergoregion, but equally-compact horizonless geometries should turn unstable when spinning sufficiently fast as they develop an ergoregion.

The ergoregion instability has been proved for rotating uniform-density stars~\cite{1978CMaPh..63..243F,Vilenkin:1978uc,CominsSchutz,1996MNRAS.282..580Y,Kokkotas:2002sf,Superradiance}, for highly-spinning boson stars~\cite{Cardoso:2007az}, and for superspinars~\cite{Cardoso:2008kj,Pani:2010jz} (i.e., string-inspired, regularized Kerr geometries spinning above the Kerr bound~\cite{Gimon:2007ur}).
The time scale of the instability depends strongly on the spin and on the compactness~\cite{Chirenti:2008pf} of the object. In particular --~because the unstable modes are those that are long-lived in the nonspinning case~-- the instability exists only for those objects which are compact enough to possess a photon sphere~\cite{Cardoso:2014sna}, i.e. their radius is smaller than the light-ring radius 
(cf.\ Ref.~\cite{Superradiance} for a detailed discussion). The photon sphere is naturally present in all models of ECOs that modify the BH geometry only at the horizon scale, for example by invoking a surface located at microscopic or even Planck distance from the would-be horizon.

The scope of this work is twofold. On the one hand, we show for the first time that the ergoregion instability generically affects ultracompact exotic objects with a perfectly reflecting surface, and might have a crucial impact on the phenomenology of these hypothetical objects. We show that the instability time scale actually \emph{increases} when the compactness of the object is extremely close to the BH limit, but this increase is only logarithmic and therefore not enough to quench the instability, even when the ECO surface is just a Planck distance from the would-be horizon.
On the other hand, we find a generic effect that can prevent the ergoregion instability from developing. Partial absorption smaller than $0.4\%$ (i.e., reflectivity only $\sim0.4\%$ smaller than unity) at the surface is sufficient to quench the instability completely. As discussed below, in an ECO this level of absorption might be naturally provided by the viscosity of the body. This finding has important consequences for the viability of ECO models.
Through this work, we use $G=c=1$ units.

\section{Setup}

\subsection{Background geometry}

We wish to describe geometries that modify the Kerr metric only at the horizon scale, as in some quantum-gravity scenarios~\cite{Mazur:2001fv,Mathur:2005zp,Skenderis:2008qn,Almheiri:2012rt,Giddings:2014ova,Holdom:2016nek,Brustein:2017kcj,Barcelo:2017lnx}. Our model is therefore very simple: we consider a geometry described by the Kerr metric
when $r>r_0$ and, at $r=r_0$, we assume the presence of a membrane with some reflective properties. Different models of ECOs are then characterized by different properties of the membrane at $r=r_0$, in particular by a (generically) frequency-dependent reflectivity. In Boyer-Lindquist coordinates, the line element at $r>r_0$ reads
\begin{eqnarray}
ds^2&&=-\left(1-\frac{2Mr}{\Sigma}\right)dt^2+\frac{\Sigma}{\Delta}dr^2-\frac{4Mr}{\Sigma}a\sin^2\theta d\phi dt   \nn \\
&+&{\Sigma}d\theta^2+
\left[(r^2+a^2)\sin^2\theta +\frac{2Mr}{\Sigma}a^2\sin^4\theta \right]d\phi^2\,,\label{Kerr}
\end{eqnarray}
where $\Sigma=r^2+a^2\cos^2\theta$ and $\Delta=r^2+a^2-2M r$. We assume that the energy and angular momentum of the membrane are negligible, so that $M$ and $J:=aM$ represent the total mass and spin of the object.

Motivated by models of microscopic corrections at the horizon scale, in the following we shall focus on the case
\begin{equation}
 r_0 = r_+ + \delta\,, \qquad 0<\delta\ll M\,,
\end{equation}
where $r_+=M+\sqrt{M^2-a^2}$ is the location of the would-be horizon. Although the above parametrization requires $a\leq M$, the latter condition is not strictly necessary. In Appendix~\ref{app:superspinars} we briefly consider violations of the Kerr bound, $a>M$, complementing previous results done in the context of the ergoregion instability of superspinars~\cite{Cardoso:2008kj,Pani:2010jz}.
Since we assume that $\delta$ is microscopic or even Planckian, it is surely much smaller than the gravitational radius of (super)massive dark objects. Clearly, the smaller $\delta$ the larger the object's compactness $M/r_0$, and $\delta\to0$ corresponds to the BH limit.
Since $r_0>r_+$, our background geometry does not have a horizon, but features an ergoregion whose outer boundary is $r_+^{\rm ER}=M+\sqrt{M^2-a^2\cos^2\theta}$. On the equatorial plane, the ergoregion extends up to $r=2M$ and therefore exists whenever $r_0<2M$.

The absence of Birkhoff's theorem in axisymmetry implies that the vacuum region outside a spinning object is not necessarily described by the Kerr geometry. For example, the spin-induced quadrupole moment of a compact star is different from the Kerr value and depends on the equation of state of the stellar interior. On the other hand, even ultracompact density stars have a maximum compactness which is anyway smaller than that of a BH, and therefore the regime $\delta\to0$ is meaningless in this case. In those ECO models that admit a BH limit, there are indications that all multipole moments of the external spacetime approach those of a Kerr BH as $\delta\to0$~\cite{Pani:2015tga}.
This intriguing property holds true at least for thin-shell gravastars~\cite{Mazur:2001fv,Visser:2003ge,Pani:2015tga,Uchikata:2015yma,Uchikata:2016qku} and
for strongly-anisotropic, incompressible neutron stars~\cite{Yagi:2015hda,Yagi:2015upa}. In fact, for $\delta\to0$, it is natural to expect that the exterior spacetime is extremely close to Kerr, unless some discontinuity occurs in the BH limit.

\subsection{Linear perturbations}

In order to study the stability of this geometry, we consider the simplest case, namely a test scalar field governed by the massless Klein-Gordon equation, $\square \Psi=0$, where the D'Alembertian operator is defined on the metric~\eqref{Kerr}. It is convenient to decompose the scalar field as follows
\begin{equation}
 \Psi(t,r,\theta,\phi)=\sum_{lm}\int d\omega e^{-i\omega t+im\phi} S_{lm}(\theta)R_{lm}(r)\,, \label{perturb}
\end{equation}
where $S_{l m}(\theta)e^{im\phi}$ are the spheroidal harmonics, which reduce to the standard scalar spherical harmonics when $a=0$ (cf., e.g., Ref.~\cite{Berti:2005gp}). With the above decomposition, the Klein-Gordon equation on the background~\eqref{Kerr} can be separated in the following system
of ordinary differential equations~\cite{Teukolsky:1972my}

\begin{eqnarray}
&&\frac{d}{dr}\left(\Delta\frac{dR_{lm}}{dr}\right)+ \left[\frac{K^{2}}{\Delta} -\lambda\right]R_{l m}=0\,,\label{wave_eq} \\
&& \frac{d}{dx}\left((1-x^2)\frac{S_{l m}}{dx}\right)+ \left[(a\omega x)^2+A_{l m}-\frac{m^2}{1-x^2}\right]S_{l m}=0\,, \nn\\\label{angular}
\end{eqnarray}
where $x\equiv\cos\theta$, $K=(r^2+a^2)\omega-am$, and the separation constants
$\lambda$ and $ A_{l m}$ are related by $\lambda \equiv  A_{l m}+a^2\omega^2-2am\omega$.
When $a=0$, the angular eigenvalues simply read $\lambda=l(l+1)$ whereas, when $a\omega\ll1$, $A_{lm}$  can be expanded as~\cite{Berti:2005gp}
\begin{equation}
  A_{lm}=\sum_{n=0}f^{(n)}_{lm}(a\omega)^{n}\,.\label{rel:eigexpans}
\end{equation}
where $f^{(n)}_{lm}$ are constant. By solving Eq.~\eqref{angular} via continued fractions~\cite{Leaver:1985ax}, we have verified that the above expansion is an excellent approximation of the exact numerical eigenvalues whenever $|a\omega|\lesssim1$ which --~as we shall see~-- is always the case for the fundamental modes when $a<M$.

It is convenient to simplify the radial equation~\eqref{wave_eq} by introducing $Y=(r^2+a^2)^{1/2}R$ and the tortoise coordinate $r_*$ such that $dr_*/dr=(r^2+a^2)/\Delta$. This yields
\begin{equation}
\frac{d^2Y}{dr_*^2}+VY=0\,, \label{final}
\end{equation}
where the frequency-dependent, effective potential reads
\begin{equation}
 V=\frac{K^2-\Delta\lambda}{(r^2+a^2)^{2}}-\frac{\Delta  \left(r \left(a^2+r^2\right) \Delta '+\left(a^2-2 r^2\right) \Delta \right)}{\left(a^2+r^2\right)^4}\,,\nn
\end{equation}
with a prime denoting a derivative with respect to $r$.

\subsection{Boundary conditions}\label{sec:BC}

After imposing physical boundary conditions at infinity and near $r\sim r_0$, Eq.~\eqref{final} defines an eigenvalue problem whose (complex) eigenvalues [the quasinormal modes (QNMs)], $\omega=\omega_R+i\,\omega_I$, are the characteristic frequencies of the system (cf., e.g., Ref~\cite{Berti:2009kk} for a recent review). In our conventions for the Fourier transform of the perturbation [cf.\ Eq.~\eqref{perturb}], a stable mode corresponds to $\omega_I<0$, whereas $\omega_I>0$ corresponds to an unstable mode with e-folding growth time scale $\tau:=1/\omega_I$.

The proper vibrations of the object require no incoming waves from infinity, i.e.~\cite{Teukolsky:1974yv}
\be
Y\sim e^{i\omega r_*}\qquad r\to\infty\,.\label{asymp sol}
\ee

The presence of a horizon would require purely ingoing waves as $r\to r_+$ (equivalently, as $r_*\to-\infty$). However, in our model some boundary condition has to be imposed at $r=r_0>r_+$. In general, the latter depends on the physical properties of the ECO surface. For the sake of generality, we shall consider several choices of boundary conditions.
The first case corresponds to a perfectly reflective surface~\cite{Saravani:2012is,Abedi:2016hgu}. For this case, we consider two sets of boundary conditions, namely
\begin{equation}
 \left\{\begin{array}{ll}
    Y(r_0)=0 & \qquad{\rm Dirichlet} \\
    dY(r_0)/dr_*=0 &\qquad {\rm Neumann} \\
   \end{array}\right.\,. \label{BCs1}
\end{equation}
In particular, the Dirichlet condition implies that the waves are totally reflected with inverted phase, whereas the Neumann boundary condition implies that they are totally reflected in phase.
Note that, imposing the Dirichlet condition on $Y$ at the surface is equivalent to imposing it directly on the Klein-Gordon field $\Psi$, whereas the same is not true for the Neumann condition. The latter imposed on $Y$ corresponds to a Robin-type boundary condition on $\Psi$.

In the case of a perfectly reflecting surface, perturbations experience an infinite potential well at $r=r_0$, so the internal geometry of the object does not play a role in the stability analysis. However, a perfectly reflecting surface is an idealization and, in reality, we expect that the surface or the ECO can absorb at least a small fraction of the radiation. The details of this case are model dependent and should be analyzed on a case-by-case analysis. However, considerable insight can be gained by parametrizing the properties of the interior through a generic reflection coefficient.
The generic solution of Eq.~\eqref{final} near $r\sim r_0$ reads
\begin{equation}
 Y(r_0)\sim A_{\rm out} e^{i k r_*(r_0)} + A_{\rm in}e^{-i k r_*(r_0)}\,,
\end{equation}
where $k^2=V(r_0)$, whereas $A_{\rm out}$ and $A_{\rm in}$ represent the amplitude of the wave reflected and incident at the surface, respectively. In order to account for a nonvanishing absorption, we shall impose a further boundary condition
\begin{equation}
 A_{\rm out} = {\cal R}\,  A_{\rm in} e^{-2i k r_*(r_0)}\,,\label{BCs2}
\end{equation}
where ${\cal R}$ is the reflection coefficient.
It is straightforward to show that Dirichlet and Neumann boundary conditions~\eqref{BCs1} correspond to ${\cal R}=-1$ and ${\cal R}=1$, respectively. As expected, both boundary conditions~\eqref{BCs1} correspond to $|{\cal R}|^2=1$, i.e. to perfect reflection.
Note also that, were $k$ real, $|{\cal R}|^2$ would precisely define the fraction of reflected scalar flux in units of the incident one at $r=r_0$~\cite{Teukolsky:1974yv}. However, in the small-$\delta$ limit, $k=\omega-m\Omega+{\cal O}(\delta/M)$ (where $\Omega=a/(2M r_+)$ is the angular velocity at the horizon of a Kerr BH) and it is therefore complex for QNMs. As we shall see, in the BH limit the imaginary part of the QNM vanishes sufficiently fast, so that $|e^{-2i k r_*(r_0)}|^2\to 1$ and therefore $|{\cal R}|^2 \approx |A_{\rm out}|^2 / |A_{\rm in}|^2$.

\subsection{Numerical procedure}

The boundary conditions~\eqref{asymp sol} and either \eqref{BCs1} or \eqref{BCs2} can be imposed on a numerical solution of Eq.~\eqref{final} by using a shooting method (cf., e.g., Ref.~\cite{Pani:2013pma} for a review in the context of BH perturbation theory). Starting with an analytical high-order series expansion at large distances, we integrate Eq.~\eqref{final} from infinity inwards up to $r=r_0$; we repeat the integration for different
values of (complex) $\omega$ until the desired boundary condition (either one of Eq.~\eqref{BCs1} or Eq.~\eqref{BCs2} with a specific value of ${\cal R}$) is satisfied. In the case of Eq.~\eqref{BCs2}, we choose the root of $k^2=V(r_0)$ such that the sign of the real part agrees with ${\rm sign}(\omega_R-m\Omega)$ as in the Kerr case. Physically, this corresponds to require that $A_{\rm out}$ and $A_{\rm in}$ are the amplitudes of the reflected and incident wave also in a reference frame co-rotating with the object.

When normalized by the mass $M$, the QNMs of the system for a given set of boundary conditions depend on two continuous, dimensionless parameters: the spin $a/M$ and the distance $\delta/M$ (or, equivalently, the compactness $M/r_0$). Furthermore, they depend on three integer numbers, namely on the angular number $l\geq0$, on the azimuthal number $m$ (such that $|m|\leq l$), and on the overtone number $n\geq0$. We shall mostly focus on the $l=m=1$ fundamental modes ($n=0$) which, in the unstable case, correspond to the modes with the largest imaginary part (i.e., with the shortest instability time scale).
In our numerical results, we also make use of the symmetry~\cite{Leaver:1985ax}
\begin{equation}
 m\to-m\,,\quad\omega_R\to-\omega_R\,,\quad A_{lm} \to A_{l-m}^*\,.\label{symmetry}
\end{equation}
The latter guarantees that, without loss of generality, we can focus on modes with $m\geq0$ only.

\section{Ergoregion instability of ECOs}

\subsection{Perfect reflection, $|{\cal R}|^2=1$}

\begin{figure}[th]
\centering
\includegraphics[width=0.47\textwidth]{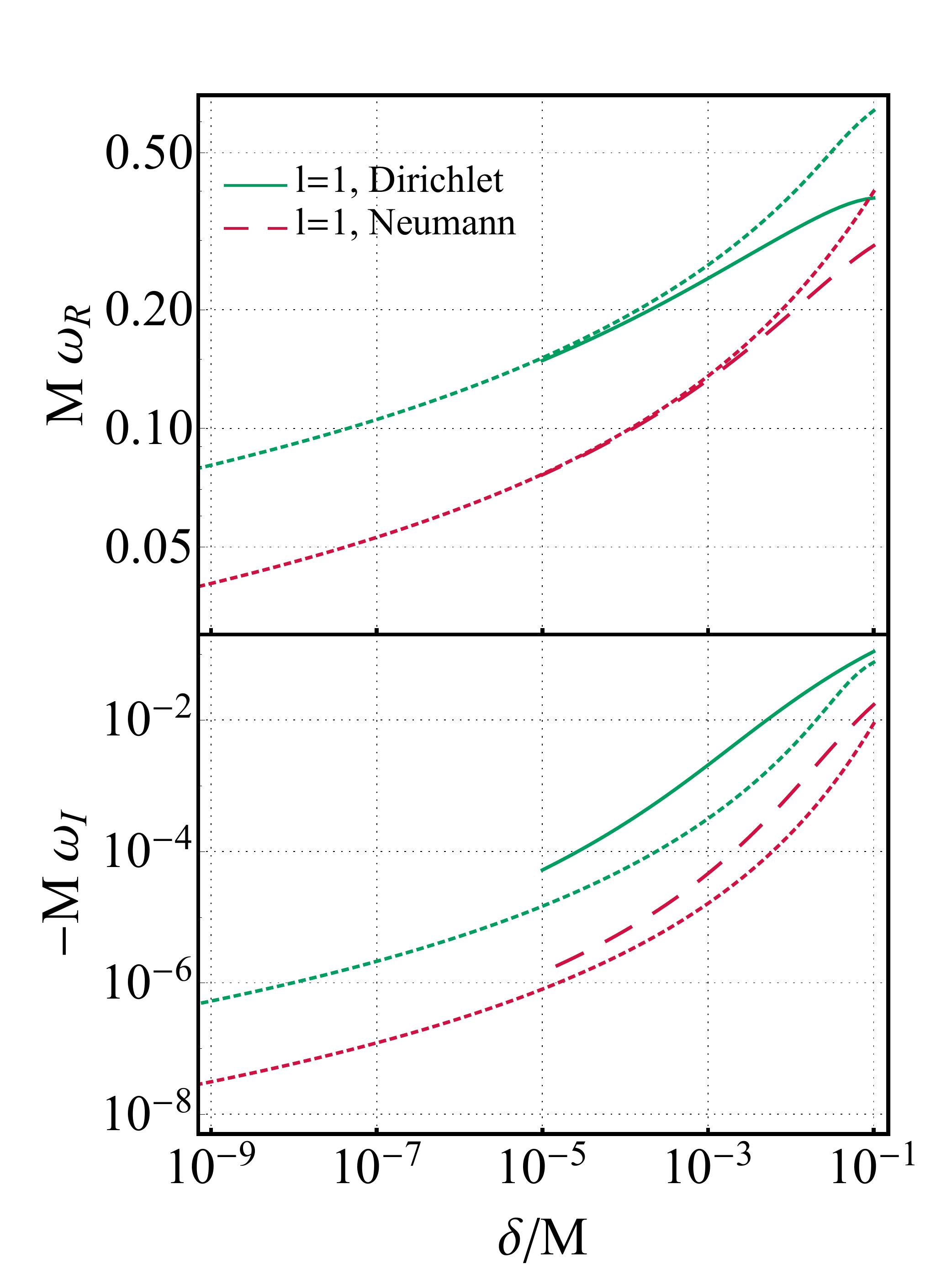}
\caption{Real (top panel) and imaginary (bottom panel) part of the fundamental QNM of an ECO with Dirichlet and Neumann boundary conditions at $r_0=2M+\delta$ as a function of the distance $\delta$ for $l=1$. The modes are stable and long lived with $\omega_I\ll \omega_R$. Small-dashed curves correspond to the analytical result derived in Appendix~\ref{app:analytics} [cf.\ Eqs.~\eqref{omegaRana0}--\eqref{omegaIana0}], which is valid for $M\omega\ll1$.
}
\label{fig:static}
\end{figure}

We start by considering the case of a perfectly reflective surface, i.e. by imposing Eq.~\eqref{BCs1} at $r=r_0$. The two choices of boundary condition correspond to two different families of modes.
In Fig.~\ref{fig:static} we show these two families for a static ECO as a function of the distance $\delta$. In the static case, the azimuthal number is degenerate and the QNMs depend only on $l$ and on the compactness.

Note that $\omega_I<0$, i.e. in the static case the modes are stable with damping time $\tau=1/|\omega_I|$. As discussed in Refs.~\cite{Cardoso:2016rao,Cardoso:2014sna}, in the BH limit ($\delta\to 0$) the QNM frequency approaches zero logarithmically\footnote{The QNM frequency is related to the inverse coordinate time that radiation takes from the light ring to the ECO surface~\cite{Cardoso:2016rao,Cardoso:2016oxy,Damour:2007ap}. In the context of GW postmerger phase, it is precisely the GW echo frequency discussed in Refs.~\cite{Cardoso:2016rao,Cardoso:2016oxy,Abedi:2016hgu}.}, whereas the damping time becomes infinite, i.e., the modes become extremely long lived.
In Appendix~\ref{app:analytics}, we derive analytical estimates for these modes in the small-frequency limit. In the nonspinning case, the analytical result reads~\cite{Vilenkin:1978uc} (see also Ref.~\cite{Cardoso:2017njb})
\begin{eqnarray}
M \omega_R&\simeq&\frac{M}{2|r_*^0|}(p\pi-\Phi)\sim |\log\epsilon|^{-1}\,, \label{omegaRana0}\\
M\omega_I &\simeq&-\beta_{l}\frac{M}{|r_*^0|}(2M\omega_R)^{2l+2}\sim -|\log\epsilon|^{-(2l+3)}\,, \label{omegaIana0}
\end{eqnarray}
where $\epsilon=\delta/(2M)$, $r_*^0=r_*(r_0)\sim 2M\log\epsilon$, $p$ is a positive odd (even) integer for Dirichlet (Neumann) boundary conditions, $\Phi$ is the phase of the wave reflected at $r=r_0$, and $\beta_{l}=\frac{(l!)^4}{[(2l)!(2l+1)!!]^2}$~\cite{Starobinsky:1973aij,Brito:2015oca}. In the small-frequency limit described in Appendix~\ref{app:analytics} we find that $\Phi\approx 3\pi$ as $\delta\to0$, for both Dirichlet and Neumann boundary conditions.

By using the definition $k^2=V(r_0)$ and the above behavior for $\omega=\omega_R+i\omega_I$, it is easy to show that $k r_*(r_0)\to0$ as $\delta\to0$, and therefore $|e^{-2i k r_*(r_0)}|^2\to 1$ as previously anticipated. In Fig.~\ref{fig:static} we compare the above estimate with our numerical results. As expected, the agreement between the analytical and the numerical results is better for the real part, since $\omega_R\gg\omega_I$. Overall the agreement is good when $M \omega \ll 1$.

The fact that the imaginary part of the modes vanishes as $\delta\to0$ suggests that these modes can turn unstable in the spinning case, due to the Zeeman splitting of the frequencies. It is straightforward to check this fact, for example by analyzing the small-spin limit. In this case, to linear order in the spin, the degeneracy of $m$ is broken and the QNM can be written as (cf., e.g., Refs.~\cite{Kojima:1992ie,Pani:2012vp,Pani:2012bp,Pani:2013pma})
\begin{equation}
 \omega_{R,I}=\omega^{(0)}_{R,I}+m(a/M)\omega^{(1)}_{R,I} +{\cal O}(a^2/M^2)\,,
\end{equation}
where $\omega^{(0)}_{R,I}$ are the (real, imaginary) part of the QNM in the static case and $\omega^{(1)}_{R,I}$ are corrections of the order unity which depend only on $l$ and on $\delta$. Since the sign of the corrections depends linearly on $m$, when $\omega^{(0)}_{I}\to0$ the ${\cal O}(a/M)$ term in the above equation can make $\omega_I>0$ (either for $m>0$ or for $m<0$) even for very small rotation rates.

\begin{figure*}[th]
\centering
\includegraphics[width=0.47\textwidth]{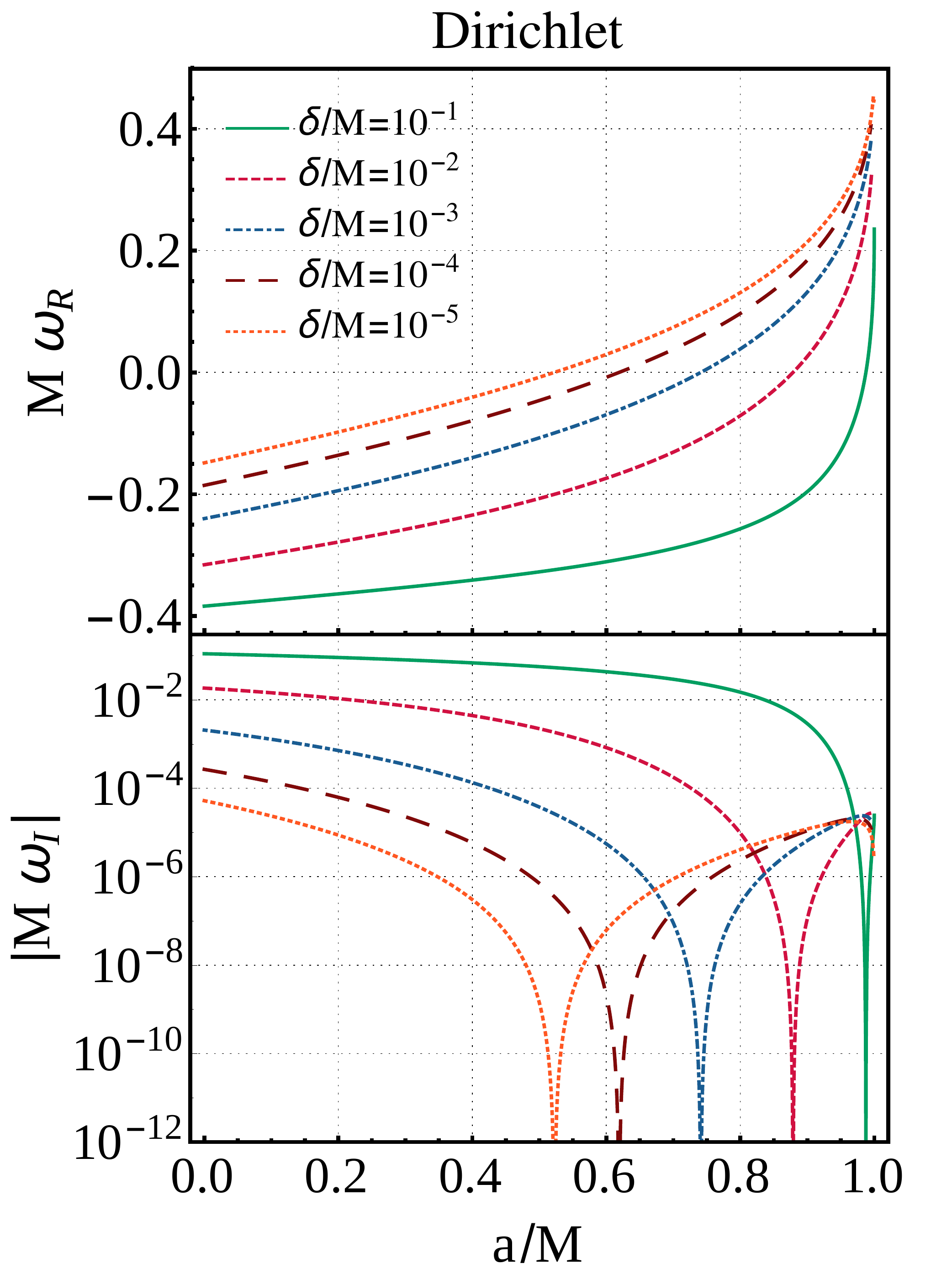}
\includegraphics[width=0.47\textwidth]{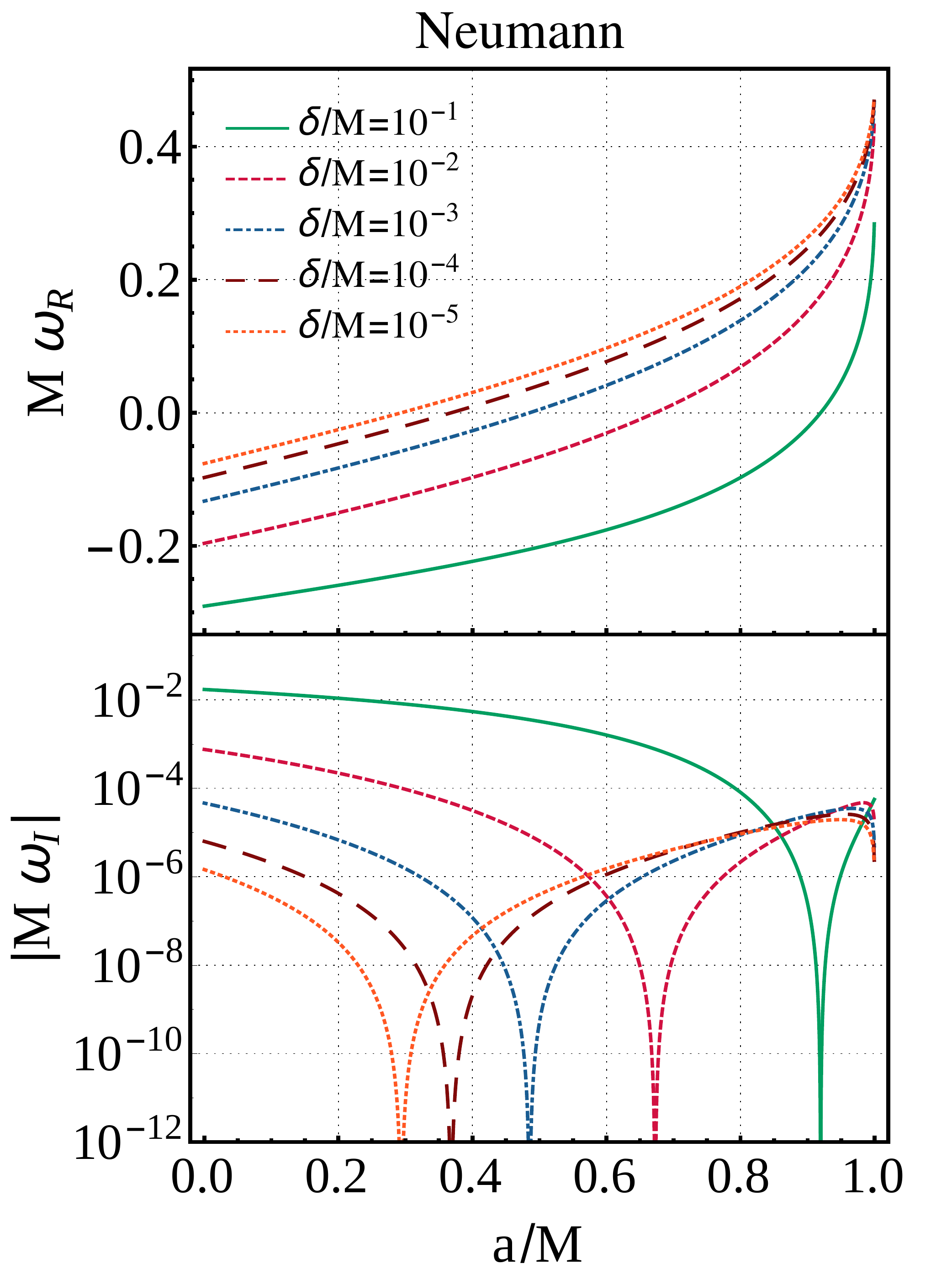}
\caption{Real (top panel) and imaginary (bottom panel) part of the fundamental QNM of an ECO as a function of the spin, for $l=m=1$ and for different values of the distance $\delta$. The left and right panels refer to Dirichlet and Neumann boundary conditions at $r_0=2M+\delta$, respectively.
The cusps in the imaginary part correspond to the threshold of the ergoregion instability.
}
\label{fig:spin}
\end{figure*}

This expectation is confirmed by our exact numerical results (which are valid for arbitrary spin) displayed in Fig.~\ref{fig:spin}. We show the fundamental modes as a function of the spin and for different values of the distance $\delta$. In Appendix~\ref{app:analytics}, we also compare the exact numerical results with the small-frequency analytical approximation (cf.\ Fig.~\ref{fig:spin_analitico}). As previously explained, by making use of the symmetry~\eqref{symmetry}, we consider both negative and positive frequencies but focus on $m\geq0$ without loss of generality.

\begin{figure*}[th]
\centering
\includegraphics[width=0.47\textwidth]{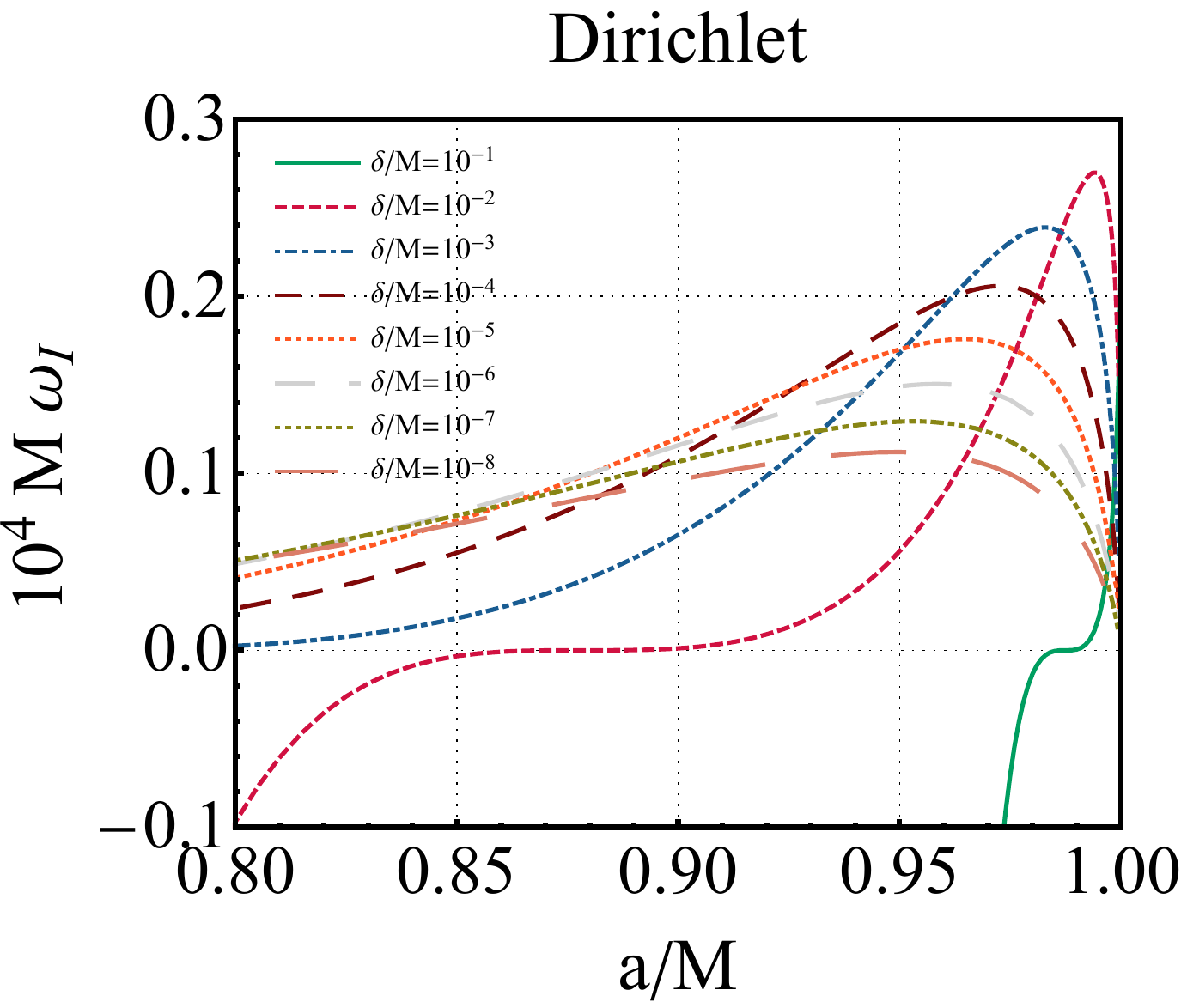}
\includegraphics[width=0.47\textwidth]{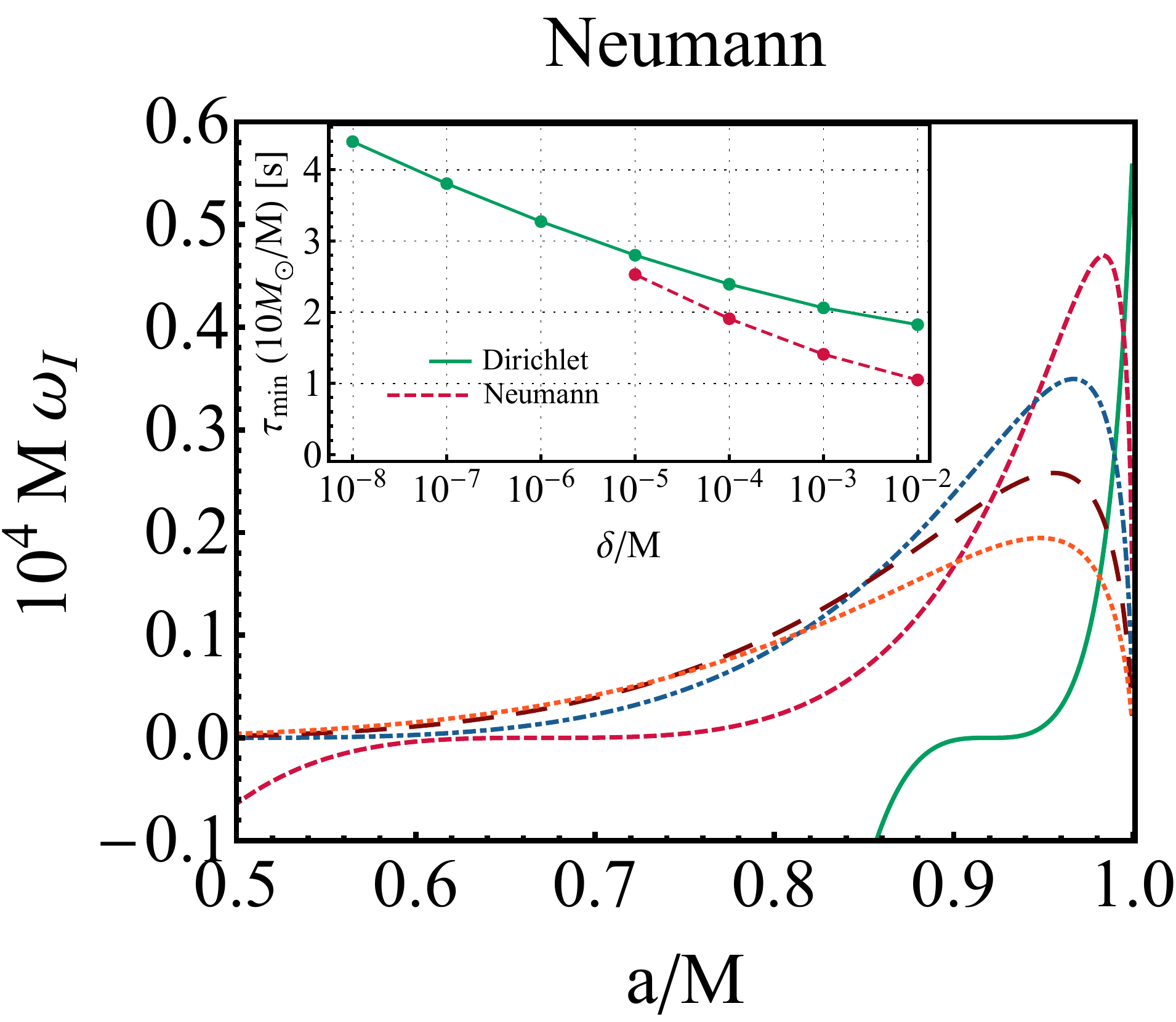}
\caption{Zoom-in version of the bottom panels of Fig.~\ref{fig:spin}. The imaginary part of the mode displays a maximum at large spin. The inset of the right panel show that the minimum instability time scale (i.e., the inverse of the maximum value of $\omega_I$) increases as $\delta\to0$ both for Dirichlet and Neumann conditions.}
\label{fig:max}
\end{figure*}

The behavior of the two families of modes is qualitatively very similar. In particular, the frequency has a zero crossing at some critical value of the spin, which monotonically decreases as $\delta\to 0$.
More importantly, also the imaginary part of the frequency changes sign for the same critical value (within our numerical accuracy). This behavior is confirmed by the analytical results derived in Appendix~\ref{app:analytics} and was also found in Ref.~\cite{Superradiance} for ultracompact spinning stars. In other words, for $a>a_{\rm crit}$ the QNMs turn from stable to unstable when $\omega_R(a_{\rm crit})=0$.\footnote{As better shown in Fig.~\ref{fig:max} for the ${\cal R}=\pm1$ cases, $\omega_I$ as a function of the spin displays an inflection point at $a=a_{\rm crit}$. This seems a generic feature of the instability, and suggests that at least a third-order expansion in the spin is needed if one wishes to properly investigate the instability in a slow-rotation approximation. We also note that this property does not hold in the partially-absorbing case, $|{\cal R}|^2<1$, discussed in the next sections.}
Note that the analytical results are valid for $M\omega\ll1$. This condition is satisfied near the critical spin $a_{\rm crit}$, whereas it is violated at large spin since $\omega_R \to m\Omega \sim m/(2 M)$ as $\delta\to0$ and $a\sim M$.

%
\begin{figure}[th]
\centering
\includegraphics[width=0.47\textwidth]{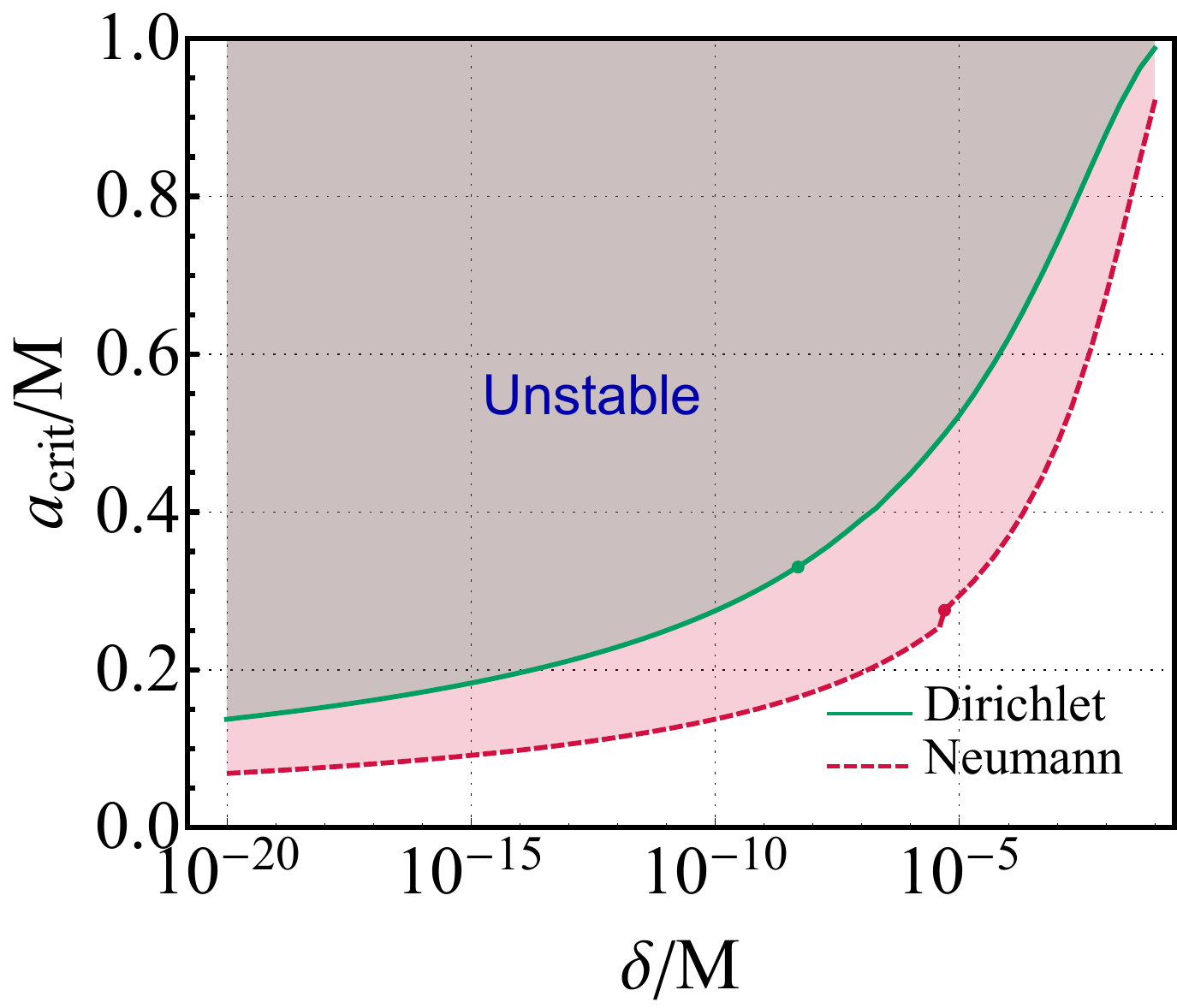}
\caption{Critical value of the spin above which an ECO with Dirichlet or Neumann boundary conditions becomes unstable against the ergoregion instability. In the Neumann case, the curve is truncated near $\delta/M\approx 10^{-5}$ due to the difficulty of tracking the modes numerically when $\delta\to0$. The circles represent the points at which the numerical result (on the right of the marker) is glued with the analytical result [cf.\ Eq.~\eqref{acrit}] valid when $\delta\to0$ (on the left of the marker).
The instability region refers to the fundamental $l=m=1$ mode and $a/M\leq1$. As discussed in Appendix~\ref{app:superspinars}, there exists an \emph{upper} critical value of $a_{\rm crit}>M$ above which the fundamental $l=m=1$ mode becomes stable again~\cite{Pani:2010jz}.
}
\label{fig:crit}
\end{figure}

As expected, the value of $a_{\rm crit}$ depends on the compactness. Figure~\ref{fig:crit} shows that $a_{\rm crit}$ decreases logarithmically as $\delta\to0$.
In the BH limit ($\delta \to 0)$, the small-frequency analytical approximation yields
\begin{equation}
 a_{\rm crit}\sim \frac{p\pi-\Phi_0}{m\log(\delta/M)}\,, \label{acrit}
\end{equation}
where $\Phi_0=\Phi(a=0)\approx 3\pi$ [cf. Eq.~\eqref{fit_phase}]. The above formula is compared to the numerical result in Fig.~\ref{fig:crit}.
%
As clear from Figs.~\ref{fig:spin} and~\ref{fig:crit}, for Neumann boundary condition the instability is more efficient. Furthermore, the instability time scale is typically short. At large spin the unstable modes have $M\omega_I\sim {\cal O}(10^{-5})$. This corresponds to an instability time scale
\begin{equation}
 \tau_{\rm instability} \sim 10^5 M\sim 5 \left(\frac{M}{10M_\odot}\right)\,{\rm s}\,,
\end{equation}
which is extremely short compared to typical astrophysical time scales. For example, for an object with $M=10\,M_\odot$, an imaginary part of the unstable QNM as small as $M\omega_I=10^{-20}$ corresponds to a time scale comparable to the Salpeter time for accretion, $\tau_{S}\sim 4.5\times 10^7\,{\rm yr}$.
On the other hand, the instability time scale is parametrically longer than the typical dynamical time scale of a compact object, the latter taken to be the light-crossing time $\sim M$.
Thus, it is unclear whether this instability can play a crucial role in the dynamics of fastly spinning ECOs, possibly preventing their existence. In the next section we shall show that this conclusion would be premature because this instability is very fragile.

Leaving aside for the moment the upcoming discussion, from the result shown in Fig.~\ref{fig:crit} we can speculate on the impact of the instability. The spin of several BH candidates as measured by X-ray observations is close to unity (cf.\ Ref.~\cite{Middleton:2015osa} for a review). For example, the spin of Cygnus X-1 (as measured through different techniques~\cite{Middleton:2015osa}) is $a/M\gtrsim0.95$. Similarly, the spin of supermassive BH candidates can be very high~\cite{Brenneman:2006hw}. From Fig.~\ref{fig:crit}, a spin close to the Kerr limit is incompatible with the ergoregion instability for any value of $\delta/M<10^{-1}$.
More robust spin measurements come from GW astronomy. The BHs formed during the coalescences detected by LIGO both have spin $a/M\sim 0.7$ with percent error at $1\sigma$ level~\cite{Abbott:2016blz,Abbott:2016nmj}. In principle, were these objects ECOs instead of BHs, according to Fig.~\ref{fig:crit} they should be unstable when $\delta/M<10^{-4}$ or $\delta/M<10^{-2}$ for Dirichlet or Neumann boundary conditions, respectively. However, it is unclear whether the ECO hypothesis can be ruled out on the basis of this instability in the postmerger phase. Indeed, it might be difficult to detect the effect of the unstable mode (whose time scale is of the order of the second or longer) on the postmerger GW signal which lasts for a much shorter time scale (of the order of tens of millisecond). The analysis of this effect goes beyond the scope of this work, and might even be superfluous on the basis of what we discuss in the next section.

\subsection{The ergoregion instability is fragile}

In this section we show that the ergoregion instability of an ECO is slightly quenched in the BH limit ($\delta\to0$) and --~most importantly~-- it is destroyed by introducing a small absorption coefficient at the surface.

\subsubsection{Large compactness (slightly) quenches the instability}

Figure~\ref{fig:max} shows a zoom-in version of the bottom panels of Fig.~\ref{fig:spin} in the high-spin regime. Interestingly, the instability is not monotonous in the spin. For a given compactness, $\omega_I$ displays a maximum at large spin.
This feature is qualitatively reproduced also by the small-frequency results, although the latter is not accurate at large spin, since $M\omega_R\sim 1/2$.
A numerical investigation of the regime $a=M$ shows that $\omega_I(a=M)$ decreases\footnote{In some cases we observe that the fundamental ($n=0$) mode becomes \emph{stable} when $a=M$, but some overtone ($n>0$) remains unstable; overall, a preliminary analysis suggests that, in general, the instability does not disappear when $a=M$.} as a function of $\delta$.
Indeed, we also observe that the height of the maximum actually \emph{decreases} when $\delta\to0$. This is better shown in the inset of the right panel of Fig.~\ref{fig:max}, which shows the minimum value of the instability time scale, $\tau_{\rm min}:=1/\omega_I^{\rm max}$, as a function of $\delta$ for both Dirichlet and Neumann boundary conditions.

In the explored range of $\delta/M$ (spanning more than $5$ orders of magnitude) the behavior of $\tau_{\rm min}$ is logarithmic and its values change only by a factor of a few. In the BH limit, the data shown in the inset of Fig.~\ref{fig:max} are fitted by
\begin{equation}
 \tau_{\rm min}\approx-\alpha\left(\frac{M}{10\,M_\odot}\right)\log(\delta/M)\, {\rm s}\,. \label{tauminfit}
\end{equation}
where approximately $\alpha\approx 0.2$ for Dirichlet boundary conditions.
Clearly, exploring the $\delta\to0$ regime is unfeasible numerically\footnote{It is technically more challenging to explore the $\delta\to0$ regime with Neumann boundary conditions. This is why the curves in Figs.~\ref{fig:crit} and \ref{fig:max} are truncated for $\delta/M\lesssim 10^{-5}$ in the Neumann case.}. Furthermore, Fig.~\ref{fig:max} shows that the maximum of the instability always occurs for large spin, where $\omega_R M\sim 1/2$ and also our small-frequency analytical approximation is unreliable.
However, the absence of any small parameter other than $\delta/M$ gives us some confidence that the results shown in the inset of Fig.~\ref{fig:max} have already converged. By (hugely!) extrapolating the fit~\eqref{tauminfit}, we obtain that even for $\delta/M\sim 10^{-40}$ the instability time scale can be as small as $20\,{\rm s}$ for a $M=10M_\odot$ object.

\begin{figure}[th]
\centering
\includegraphics[width=0.47\textwidth]{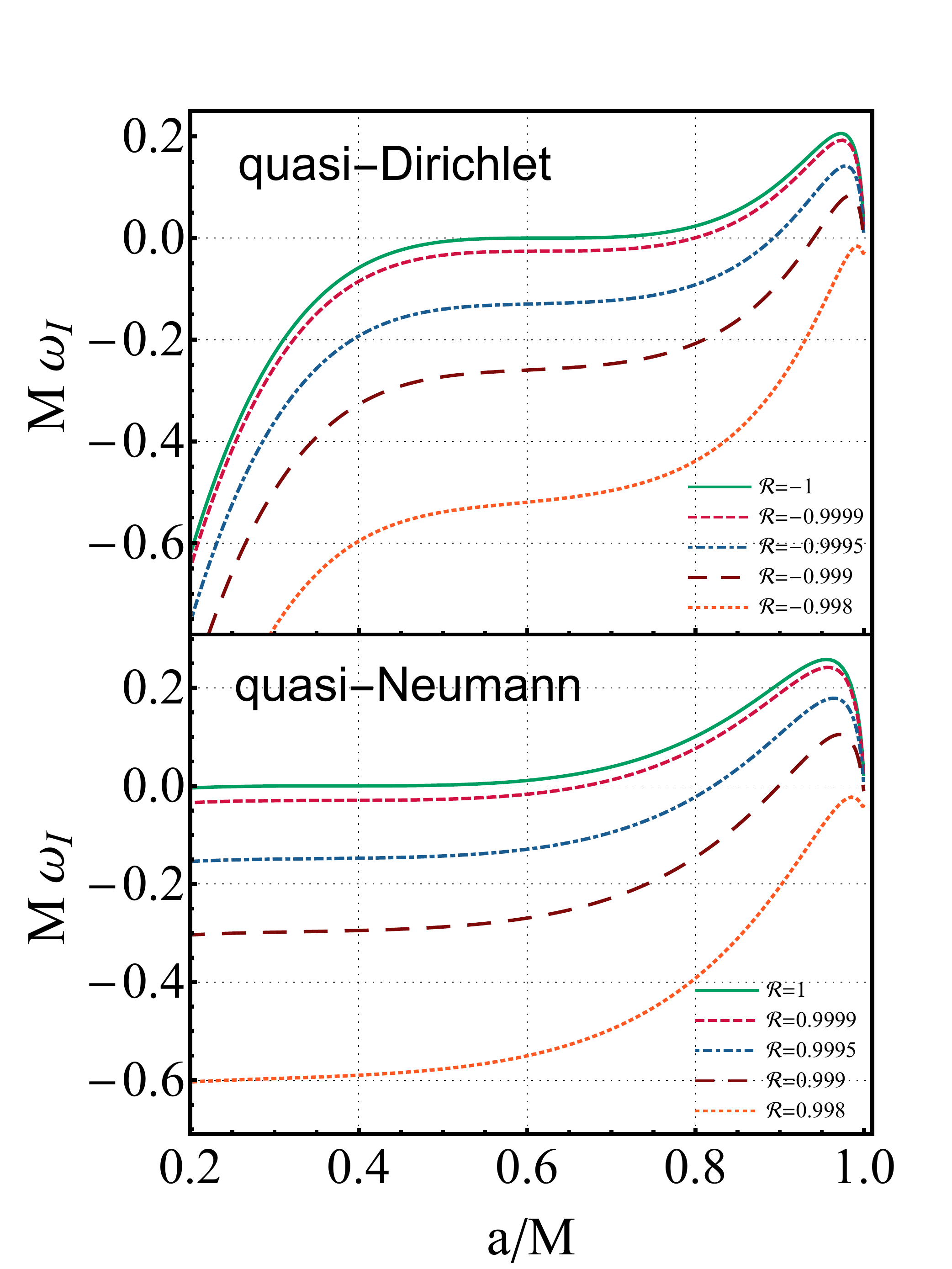}
\caption{Same as Fig.~\ref{fig:max} but for $\delta=10^{-4} M$ and different reflectivity coefficients. The top and bottom panels refer to quasi-Dirichlet (${\cal R}\gtrsim-1$) and quasi-Neumann (${\cal R}\lesssim 1$) boundary conditions. In both cases, when $|{\cal R}|^2\lesssim 0.996$, the instability disappears.}
\label{fig:reflectivity}
\end{figure}

\subsubsection{Partial absorption ($|{\cal R}|^2<1$) destroys the instability}

The case of a perfectly reflecting surface is an idealization which can never be realized in nature. In reality, we expect a compact object to absorb part of the radiation (e.g., through fluid mode excitations, dissipation, viscosity, nonlinear effects, etc...). Given the fact that Kerr BHs absorb efficiently the negative-energy modes that might form inside their ergoregion, it is relevant to ask whether some absorption at the ECO surface can quench the instability previously discussed.

This is indeed the case, as shown in Fig.~\ref{fig:reflectivity}, where we have considered small deformations near perfect reflectivity, dubbed ``quasi-Dirichlet'' and ``quasi-Neumann'' cases. The former is realized when ${\cal R}\gtrsim -1$, whereas the latter is realized when ${\cal R}\lesssim 1$. Clearly, both cases are characterized by $|{\cal R}|^2\lesssim1$ which, from the discussion of Sec.~\ref{sec:BC}, implies that a fraction $1-{\cal R}^2$ of the energy flux incident near $r=r_0$ is absorbed. Note that ${\cal R}$ can be any complex number such that $|{\cal R}|^2\leq1$. For simplicity, we consider it to be real and frequency independent, but we have checked that including a phase does not change qualitatively our results.

Although not shown, the real part of the fundamental QNM is unaffected by different values of ${\cal R}$ (at least as long as $|{\cal R}|^2\approx1$ and ${\cal R}$ is frequency independent), but their imaginary part changes drastically\footnote{Incidentally, this behavior is akin to the one observed for the QNMs of a small BH in anti de Sitter space~\cite{Cardoso:2004hs,Superradiance}. The real part of the BH QNMs is the same as that of the normal modes of anti de Sitter, whereas the presence of small absorption (provided by the small BH horizon) introduces a small imaginary part.}. Indeed, for a given compactness, the critical value $a_{\rm crit}$ \emph{increases} as $|{\cal R}|^2$ decreases, up to a point at which the fundamental mode becomes stable. Figure~\ref{fig:reflectivity} shows that for $\delta/M=10^{-4}$ the instability disappears completely when $|{\cal R}|^2\lesssim 0.996$, both for quasi-Dirichlet and for quasi-Neumann conditions. The critical value of $|{\cal R}|^2$ is only mildly dependent on $\delta$, at least in the range $\delta/M\in(10^{-8},10^{-3})$.
Although we anticipated this result, it is interesting that the amount of absorption needed to totally quench the instability is rather small. Even for the most unstable modes, an absorption rate at the level of $0.4\%$ is sufficient to quench the ergoregion-instability mode completely.

A natural question is whether this level of absorption is achievable for an ECO. The properties of an ECO depend on the specific model, but should generically be even more extreme than those of an ordinary neutron star. For neutron stars, the most efficient absorption mechanism is due to viscosity. A rough estimate of the kinematic viscosity yields~\cite{1987ApJ...314..234C}
\begin{equation}
 \nu \approx 10^{-17}\left(\frac{\rho}{10^{14}\,{\rm g/cm}^3}\right)^{5/4}\left(\frac{T}{10^8\, K}\right)^{-2}\, {\rm s}\,. \label{kinviscosity}
\end{equation}
where $\rho$ and $T$ are the typical density and temperature of a neutron star, respectively.
As a response of some external perturbation, a viscous fluid can dissipate radiation. The fraction of gravitational energy converted into mechanical energy in a viscous, compressible fluid was estimated in Refs.~\cite{1971ApJ...165..165E,1985ApJ...292..330P}, finding that the dissipation occurs through sound waves which propagate into the interior of the fluid and through shear waves which heat the surface. In the limit $\nu\omega\ll1$ which is valid in the entire parameter space of interest, and after an angle average, the fraction of absorbed energy in the flat spacetime approximation reads~\cite{1971ApJ...165..165E}
\begin{eqnarray}
 e&\sim&\frac{64 \rho}{3\omega^2}(\omega\nu)^{3/2} \label{fraction}\\
 &\approx& 0.004\, \left(\frac{M}{r_0}\right)^{27/4} \left[\frac{10^3\, K}{T}\right]^{3} \sqrt{\frac{0.01}{\omega M}}\left(\frac{20 M_\odot}{M}\right)^4\,,\nn
\end{eqnarray}
where in the last step we have normalized the physical quantities by their typical values expected for an ECO in the BH limit, namely density similar to that of a fastly spinning Kerr BH, $r_0\sim M$, and a low temperature.
As a reference, the local temperature of an isolated gravastar is of the order of the Hawking temperature $T\sim \hbar/(k_B M)\approx 10^{-7}\,{\rm K}$ for $M\sim 20\,{\rm M_\odot}$, even in the absence of a horizon~\cite{Mazur:2001fv}. This temperature is negligible in realistic astrophysical scenarios, and in practice the object would be in thermal equilibrium with its much hotter environment. The temperature of the interstellar medium typically ranges between $10\,{\rm K}$ and $10^4\,{\rm K}$, so the normalization $T\approx 10^3\,{\rm K}$ adopted above might be considered as a conservative upper bound.
Overall, the fraction of absorbed energy depends strongly on the mass and the temperature, and it is therefore model dependent; the estimate in Eq.~\eqref{fraction} is only indicative and shows that absorption at percent level can be naturally achieved by ECOs.
%

\section{Conclusion} \label{sec:conclusion}
Motivated by recent studies on quantum-gravity phenomenology at the horizon scale, we studied the stability of a model of ultracompact exotic object, consisting in a Kerr geometry excised at $r\leq r_0$. We studied test scalar perturbations on this geometry and imposed several boundary conditions at $r_0=r_++\delta$ (with $\delta\ll M$), which should account for different properties of the ECO surface. We found two interesting properties:
\begin{enumerate}
 \item For a perfectly reflecting surface, the model generically suffers from the ergoregion instability at rotation rates above a critical value. This critical value depends on the compactness and decreases logarithmically as $\delta\to0$. The instability time scale slightly increases as $\delta\to0$, but only logarithmically. For perfectly reflecting boundary conditions, the ergoregion instability is much more effective than accretion but its time scale is parametrically longer than the dynamical time scale of compact objects. Overall, in this idealized case the instability might play an important role in the dynamics of fastly-spinning ECOs.
 \item The situation changes dramatically when the surface is not perfectly reflective. A small absorption at the level of $\sim 0.4\%$ (corresponding to a reflectivity coefficient $|{\cal R}|^2\sim0.996$) is sufficient to destroy the instability even in the most unstable cases and for any value of the spin. Moderately spinning objects need even smaller absorption rates to remain stable against the ergoregion instability.
\end{enumerate}

Altogether, our results suggest that fastly-spinning ECOs with perfectly reflective boundaries might not be viable objects but, at the same time, we expect any realistic model to introduce some absorption, if only because of the interaction with the exotic material of the body.
Whether this is enough to quench the instability is a question that can be analyzed for some specific model, but a simple estimate based on the absorption of GWs by a neutron star suggests that subpercent absorption can be naturally achieved in ECOs [cf.\ Eq.~\eqref{fraction}]. Furthermore, because the amount of absorption which seems to be required to quench the instability is small, we expect that previous studies based on perfectly reflecting objects (e.g.\ concerning GW echoes~\cite{Cardoso:2016rao,Cardoso:2016oxy,Abedi:2016hgu} or the tidal Love numbers~\cite{Cardoso:2017cfl}) would remain qualitatively (and probably also quantitatively) correct even in the presence of absorption. Indeed, an absorption rate smaller than percent level --~as required for the model to allow for stable rotating solutions~-- will not change significantly the GW phenomenology of these objects. An interesting possibility that is worth studying is whether the amplitude of the GW echoes~\cite{Cardoso:2016rao,Cardoso:2016oxy,Abedi:2016hgu} for spinning objects is amplified by the ergoregion instability discussed here.

We have mostly focused on $l=m=1$ fundamental modes, but our results are qualitatively valid also for higher values of $(l,m)$. However, the instability is exponentially suppressed in the eikonal limit~\cite{CominsSchutz,1996MNRAS.282..580Y,Cardoso:2014sna} so higher-$l$ modes are less phenomenologically relevant. Likewise, we have checked that overtones $(n>0$) behave similarly to the fundamental mode\footnote{It is interesting to note that --~in some corners of the parameters space~-- some overtone and some $l>1$ modes can have imaginary part which is larger (in the unstable regime) or smaller (in the stable regime) relative to the fundamental mode. When this happens the linear dynamics will be effectively governed by these modes, with properties similar to those shown in the main text.} ($n=0$).

The most natural extension of this work is the study of gravitational perturbations. This case is considerably more involved because of the interaction with the material of the object and because the boundary conditions are less trivial. At the same time, this extension is particularly important in order to quantify the amount of absorption that is needed to quench the ergoregion instability of gravitational perturbations, and to fully assess the viability of spinning ECOs.
In this respect, it is tantalizing to note that an absorption rate of $\sim 0.4\%$ corresponds to the maximum superradiant amplification factor for scalar perturbations of a Kerr BH~\cite{Press:1972zz,Superradiance}. If this analogy extends to the electromagnetic and gravitational cases, a much higher absorption rate would be required to quench the instability, although we expect the result will crucially depend on the spin and on the compactness of the object.

Our results are valid for those models which predict almost perfect reflection at the surface~\cite{Saravani:2012is,Abedi:2016hgu}, but our framework can be applied directly to other specific models of ECOs, for example wormholes and gravastars (cf.\ Ref.~\cite{Cardoso:2017njb} for a review). In these models, the reflection coefficient is generically frequency dependent, as recently shown in Ref.~\cite{Mark:2017dnq} in the case of a static wormhole. 
A natural application and a reliability test of our work can be done by
computing ${\cal R}(\omega)$ for specific models of spinning ECOs and mapping it to our
framework.

It would also be interesting to extend our analytical estimates to the highly-spinning case, using the same approximation adopted in Refs.~\cite{Starobinsky:1973aij,Hod:2017vld}.
The study of more generic boundary conditions (e.g.\ of Robin type) is also left for future work.

Finally, although our results suggest that the ergoregion instability might be avoided in certain models of ECOs, understanding its evolution and end point (likely a slowly-spinning ECO at the threshold of the instability, $a\approx a_{\rm crit}$), as well as studying the evolution of other possible nonlinear instabilities of ECOs~\cite{Cardoso:2014sna} remain interesting open problems.

\begin{acknowledgments}
We thank Leonardo Gualtieri for interesting discussions, Vitor Cardoso for relevant comments on the draft and for suggesting the correspondence between the threshold absorption coefficient and the maximum superradiant amplification, and Sam Dolan for interesting comments.
PP acknowledges financial support provided under the European Union's H2020 ERC, Starting Grant agreement no.~DarkGRA--757480.
This project has received funding from the European Union's Horizon 2020 research and innovation programme under the Marie Sklodowska-Curie grant agreement No 690904, the ``NewCompstar'' COST action MP1304, and from FCT-Portugal through the projects IF/00293/2013.
\end{acknowledgments}
%

\appendix

\section{QNMs of ECOs in the small-frequency limit} \label{app:analytics}

In this Appendix we derive analytical expressions for the QNMs of a spinning ECO in the small-frequency regime. The derivation is a straightforward extension of previous work by Vilenkin~\cite{Vilenkin:1978uc} and by Starobinski~\cite{Starobinsky:1973aij} (cf.\ also Ref.~\cite{Cardoso:2008kj}) and it is based on a matched asymptotic expansion.

At infinity, the radial wave equation~\eqref{wave_eq} reduces to
\begin{equation}
\partial_{r}^2\left(r R_{lm}\right)+\left[\omega^2-l(l+1)/r^2\right](rR_{lm})=0 \,, \label{wave_eq_inf}
\end{equation}
whose general solution is a linear combination of Bessel functions. The requirement of purely outgoing waves at infinity fixes the ratio of the two free coefficients. When $r\omega\ll1$, such solution reads
\begin{equation}
R_{lm} \sim \alpha \frac{(\omega/2)^{l+1/2}}{\Gamma(l+3/2)} r^l + \beta \frac{(\omega/2)^{-l-1/2}}{\Gamma(-l+1/2)} r^{-l-1} \ , \label{sol_wave_eq_inf}
\end{equation}
where the absence of ingoing waves at infinity implies $\beta=-i\alpha e^{i \pi l}$.

On the other hand, in the region near the compact object the radial wave equation~\eqref{wave_eq} reduces to
\begin{equation}
\Delta \partial_{r} \left(\Delta \partial_{r} R_{lm}\right)+\left[r_{+}^4 \left(\omega-m \Omega\right)^2 -l (l+1) \Delta \right] R_{lm} = 0 \ . \label{wave_eq_hor}
\end{equation}
The general solution of Eq.~\eqref{wave_eq_hor} is a linear combination of hypergeometric functions. The ratio of the two free constants is fixed by the boundary conditions imposed at $r=r_0$. The large-\emph{r} behavior of the solution is
\begin{eqnarray}
\nonumber R_{lm} &\sim& \left(\frac{r}{r_{+}-r_{-}}\right)^l \frac{\Gamma(2l+1)}{\Gamma(l+1)} \left[A \ \frac{\Gamma(1-2 i \varpi)}{\Gamma(l+1-2 i \varpi)} \ + \right.\\
\nonumber &+& \left. B \ \frac{\Gamma(1+2 i \varpi)}{\Gamma(l+1+2 i \varpi)} \right] + \\
\nonumber &+& \left(\frac{r}{r_{+}-r_{-}}\right)^{-l-1} \frac{\Gamma(-2l-1)}{\Gamma(-l)} \left[A \ \frac{\Gamma(1-2 i \varpi)}{\Gamma(-l-2 i \varpi)} \ + \right. \\
&+& \left. B \ \frac{\Gamma(1+2 i \varpi)}{\Gamma(-l+2 i \varpi)} \right] \ , \label{sol_wave_eq_hor}
\end{eqnarray}
where $\varpi \equiv \left(\omega-m \Omega \right) r_{+}^2/(r_{+}-r_{-})$, $r_-=M-\sqrt{M^2-a^2}$, and the ratio $B/A$ is fixed by the boundary conditions.

The matching of the Eq.~\eqref{sol_wave_eq_inf} and Eq.~\eqref{sol_wave_eq_hor} in the intermediate region yields~\cite{Cardoso:2008kj}
\begin{equation}
\frac{B}{A} = - \frac{R_{-}+i (-1)^l (\omega(r_{+}-r_{-}))^{2l+1} L S_{-}}{R_{+}+i (-1)^l (\omega(r_{+}-r_{-}))^{2l+1} L S_{+}} \ , \label{B_A}
\end{equation}
where
\begin{equation}
L=\frac{\pi}{2^{2l+2}} \frac{(\Gamma(l+1))^2}{\Gamma(l+3/2) \Gamma(2l+2) \Gamma(2l+1) \Gamma(l+1/2)}
\end{equation}
and $R_{\pm} = \frac{\Gamma(1\pm 2 i \varpi)}{\Gamma(l+1\pm 2 i \varpi)}$, $S_{\pm} = \frac{\Gamma(1\pm2 i \varpi)}{\Gamma(-l\pm2 i \varpi)}$.

\begin{figure}[th]
\centering
\includegraphics[width=0.47\textwidth]{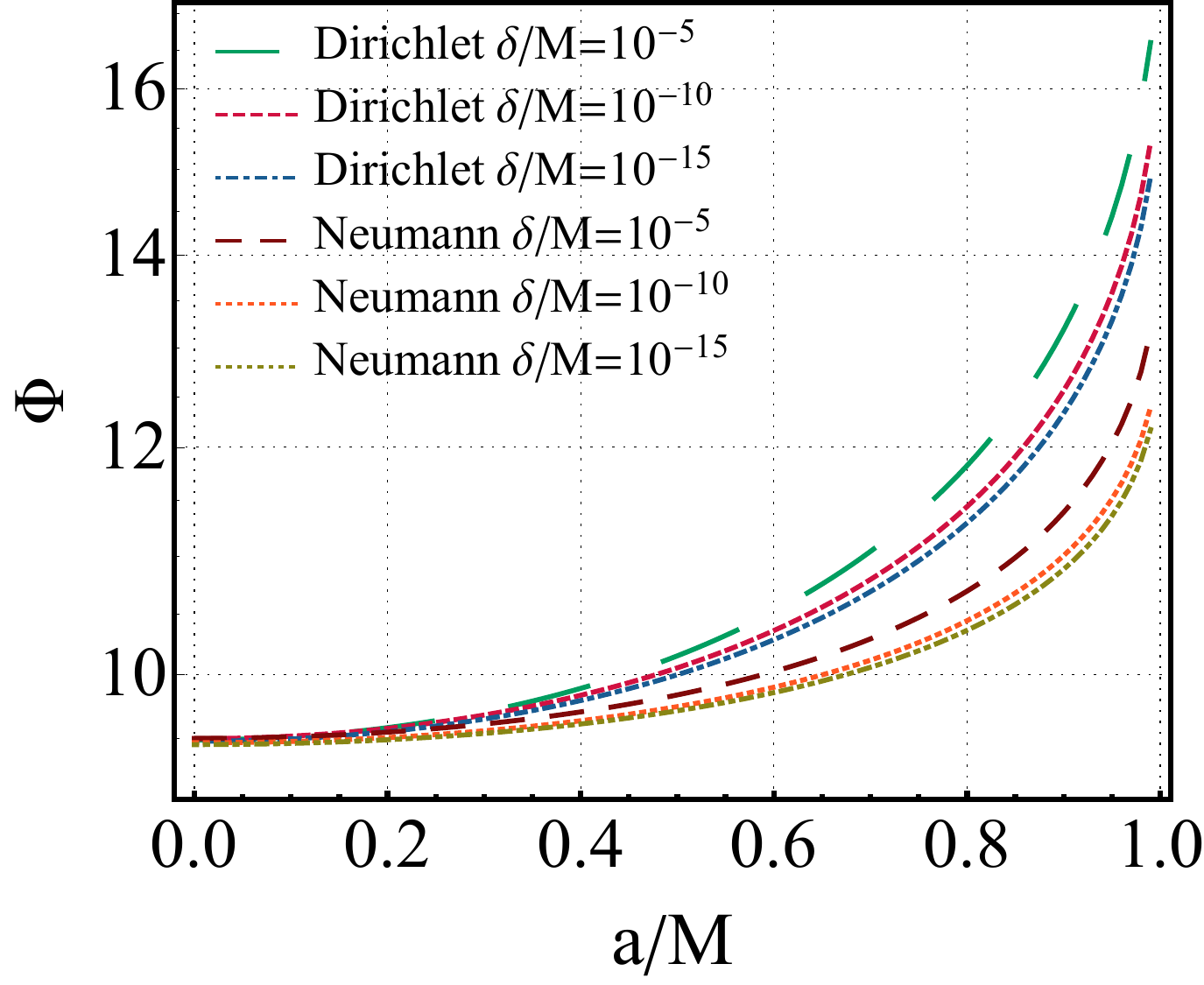}
\caption{Phase of the reflected wave as a function of the spin, both for Dirichlet and Neumann boundary conditions and for different values of the distance $\delta$. In the nonrotating case, $\Phi\approx 3\pi$ for both choices of boundary conditions.}
\label{fig:phase_appendix}
\end{figure}

\begin{figure*}[th]
\centering
\includegraphics[width=0.47\textwidth]{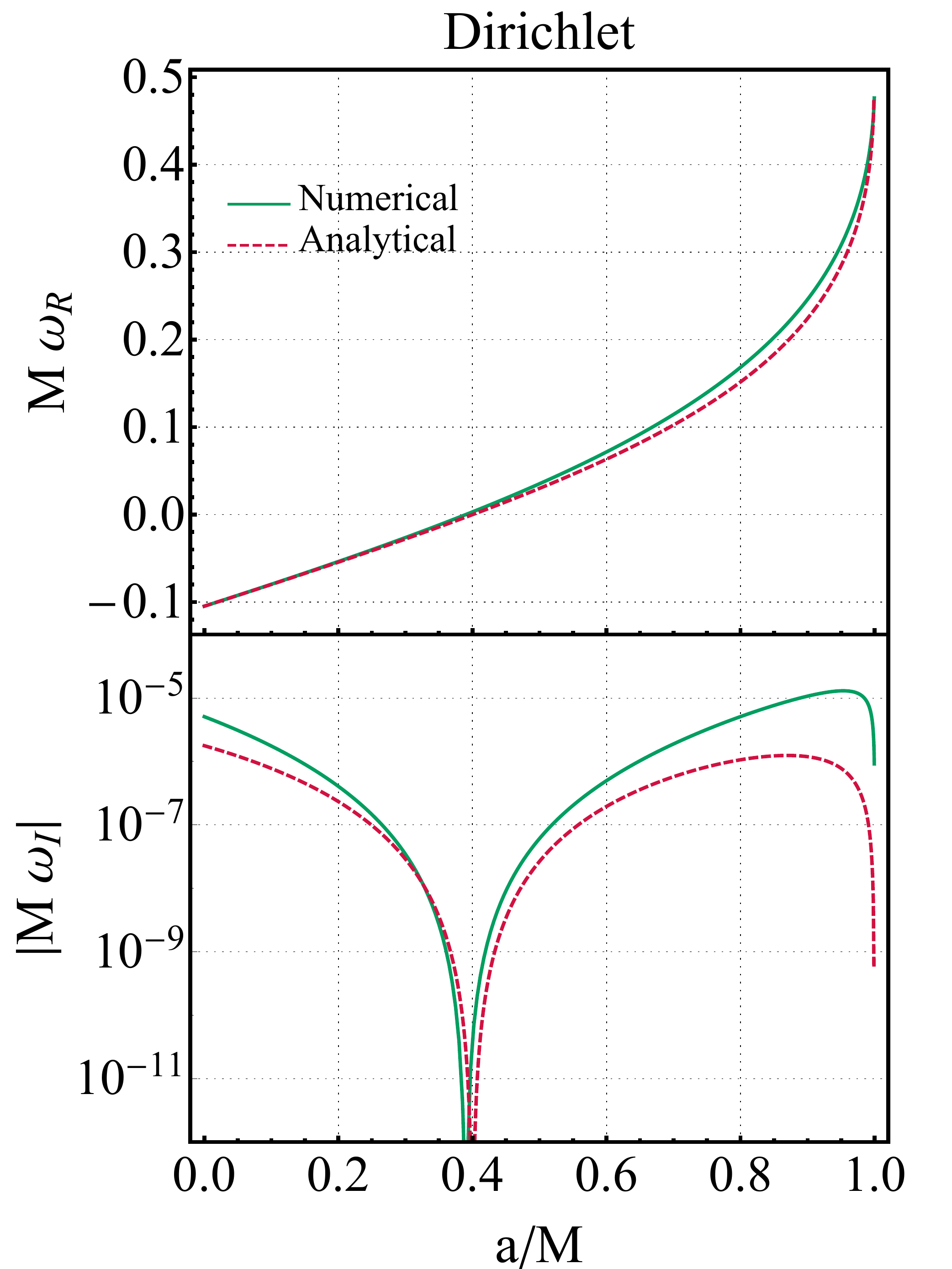}
\includegraphics[width=0.47\textwidth]{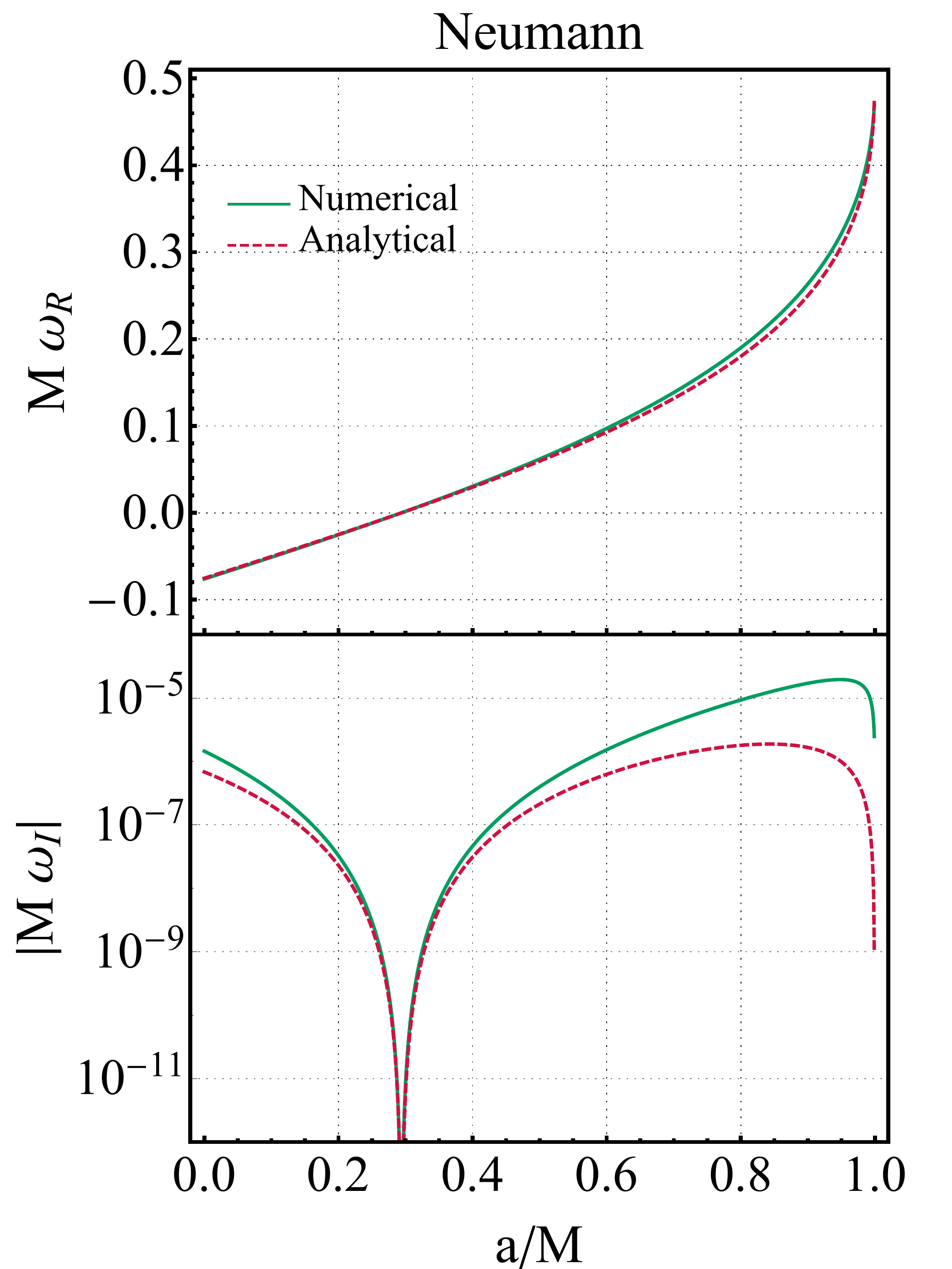}
\caption{Real (top panel) and imaginary (bottom panel) part of the fundamental QNM of an ECO as a function of the spin, for $l=m=1$. The left (right) panels refer to Dirichlet (Neumann) boundary conditions at $r_0=2M+\delta$ with $\delta/M=10^{-7}$ ($\delta/M=10^{-5}$).
Dashed curves correspond to the analytical result [cf. Eqs.~\eqref{omegaR_rot}-\eqref{omegaI_rot}], which is valid for $M \omega \ll 1 $.
}
\label{fig:spin_analitico}
\end{figure*}

Once the desired boundary conditions (either the Dirichlet or the Neumann) are imposed at $r=r_0$, Eq.~\eqref{B_A} becomes an algebraic equations for the complex frequency $\omega$ in terms of $(l,m)$, $r_0$ and the spin $a$. When $\delta\to0$, the QNMs in the small-frequency regime can be written as~\cite{Vilenkin:1978uc,Starobinsky:1973aij}
\begin{eqnarray}
\omega_{R} &\simeq& \frac{1}{2|r_*^0|}(p\pi-\Phi)+ m \Omega\,, \label{omegaR_rot}\\ 
\omega_{I} &\simeq& 2\beta_l \frac{r_{+} M}{|r_*^0|} \left[\frac{m \Omega-\omega_{R}}{r_{+}-r_{-}}\right] [\omega_{R}(r_{+}-r_{-})]^{2l+1}\,,  \label{omegaI_rot}
\end{eqnarray}
where $r_*^0$ is the tortoise coordinate of the Kerr metric evaluated at $r=r_0$. The above equations reduce to Eqs.~\eqref{omegaRana0}-\eqref{omegaIana0} in the nonrotating case. (For highly-spinning objects the result is more complex and was recently derived in Ref.~\cite{Hod:2017cga}.) Since $\omega_I\ll\omega_R$ in the $\delta\to0$ limit, the phase $\Phi$ can be computed by solving the algebraic equation Eq.~\eqref{B_A} with $\omega=\omega_R$, either with Dirichlet or Neumann boundary conditions on $Y$ (defined in terms of $R_{lm}$ above Eq.~\eqref{final}).
Figure~\ref{fig:phase_appendix} shows the phase for different values of the distance $\delta$ spanning over ten orders of magnitude. As $\delta \rightarrow 0$, the phase approaches a constant value which is well fitted by
\begin{equation}
\Phi = \frac{\Phi_0}{(1-\chi^2)^B} \left(1+ C_1 \chi + C_2 \chi^2 + C_3 \chi^3 + C_4 \chi^4\right)\,, \label{fit_phase}
\end{equation}
where $\chi=a/M$ and $\Phi_0 \approx (9.47,9.45)$, $B \approx (1.1,1.06)$, $C_1 \approx (0.03,0.01)$, $C_2 \approx (-1.1, -1)$, $C_3 \approx (0.36,0.16)$, and $C_4 \approx (-0.31,-0.13)$ for Dirichlet or Neumann boundary conditions, respectively. By replacing Eq.~\eqref{fit_phase} into Eqs.~\eqref{omegaR_rot} and \eqref{omegaI_rot}, we get an analytical estimate of the QNMs of an ECO in the small-frequency regime. Figure~\ref{fig:spin_analitico} shows a comparison between the numerical and the analytical results both for Dirichlet and Neumann boundary conditions at $r=r_0$. As in the nonspinning case, the agreement between the analytical and the numerical results is better for the real part, and it is overall good when $M \omega \ll 1$.
%

\section{Ergoregion instability of superspinars ($a/M>1$)} \label{app:superspinars}

For completeness, in this appendix we briefly consider scalar perturbations of an ECO violating the Kerr bound, namely with $a/M>1$. This case was previously studied in the context of the ergoregion instability of superspinars~\cite{Cardoso:2008kj,Pani:2010jz}. When $a/M>1$, the geometry does not possess an event horizon so $r_0$ is arbitrary. However, to avoid naked singularities and closed-timelike curves, the condition $r_0>0$ should be enforced~\cite{Pani:2010jz}. 

\begin{figure*}[th]
\centering
\includegraphics[height=0.37\textwidth]{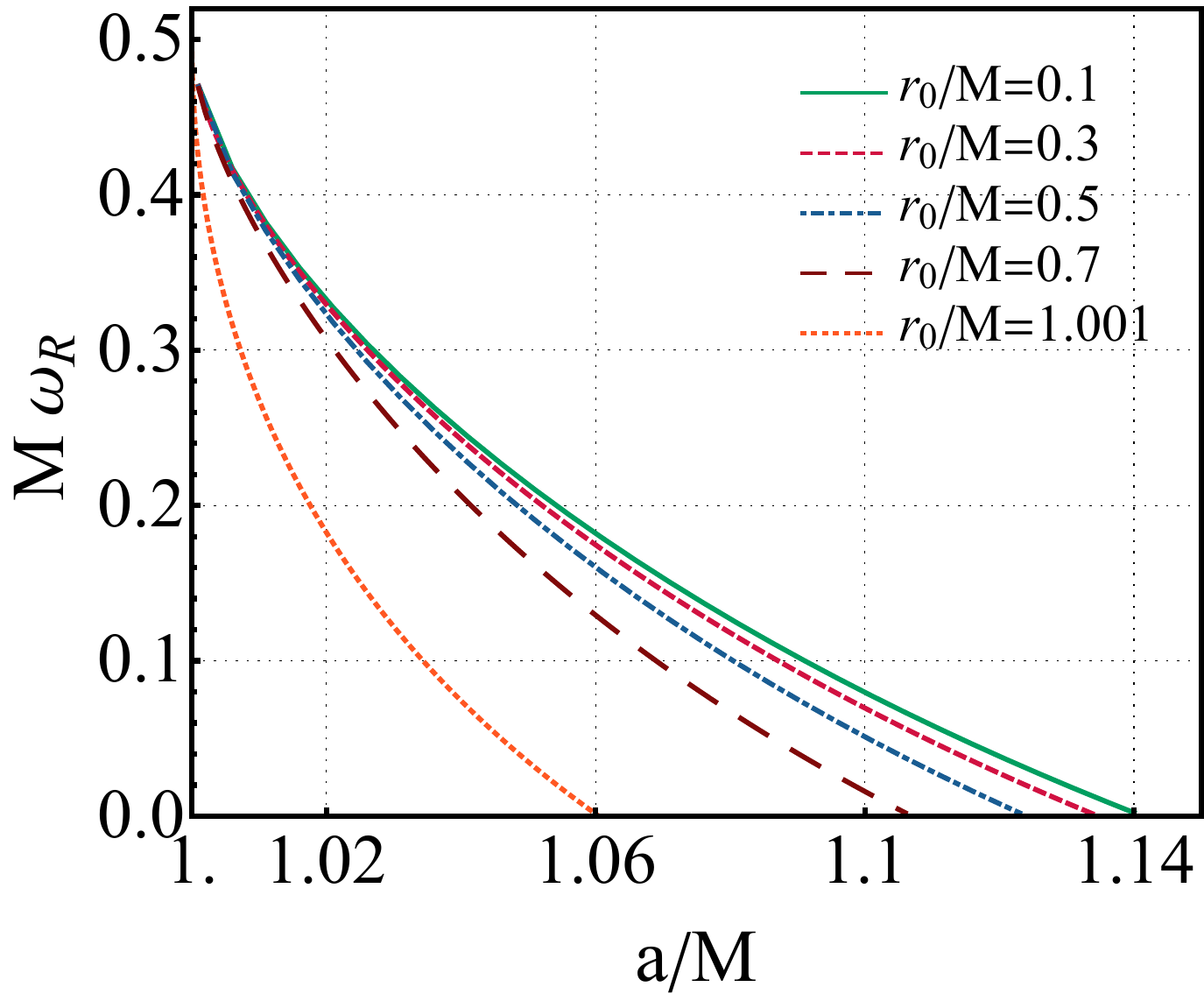}
\includegraphics[height=0.37\textwidth]{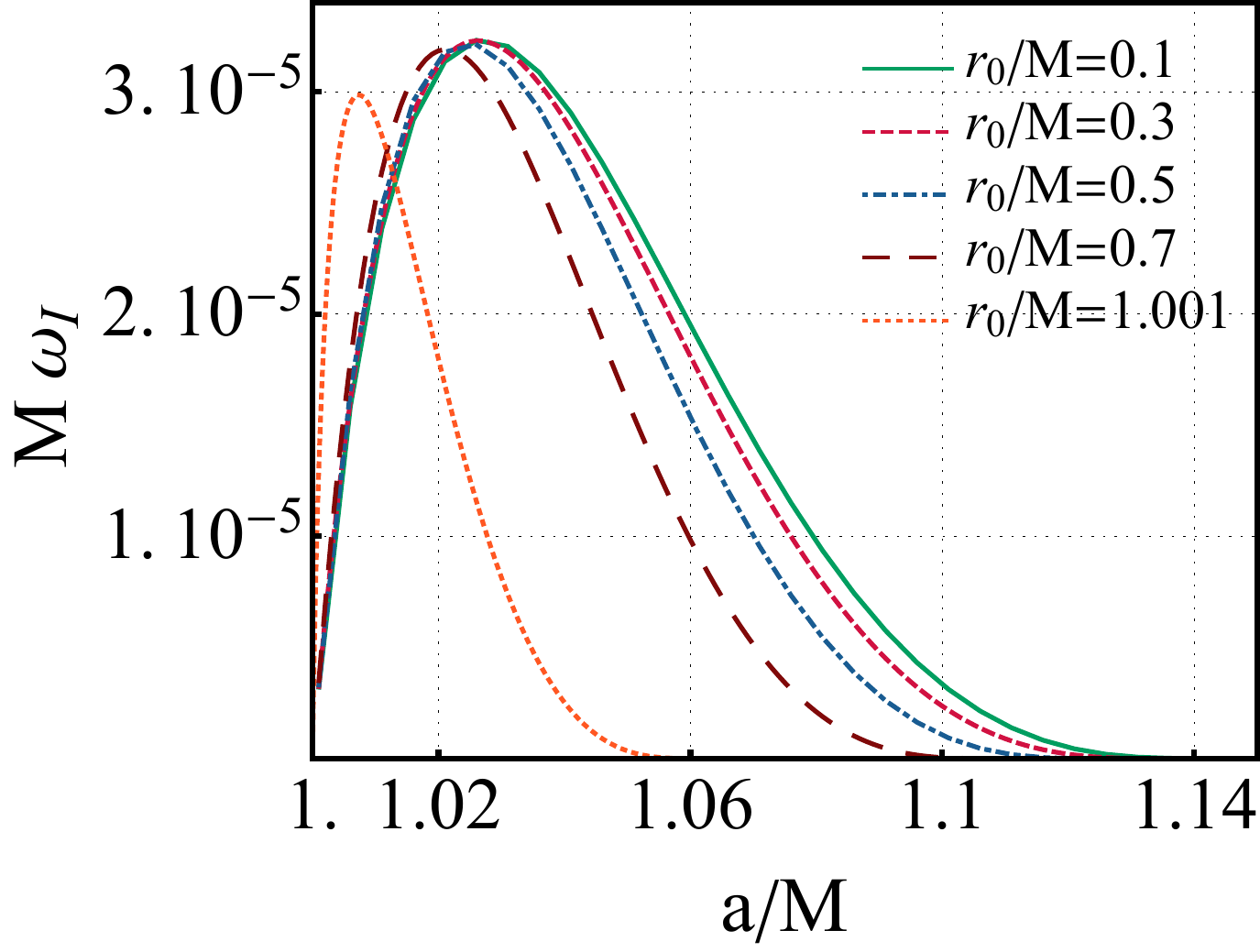}
\caption{Real (left) and imaginary (right) part of the fundamental $l=m=1$ mode for a model of superspinning ($a>M$) ECO with Dirichlet boundary conditions imposed at $r=r_0$ for different choices of $r_0$. The mode is unstable (i.e., $\omega_I>0$) only for $a<a_{\rm crit}(r_0)$. The dependence of the function $a_{\rm crit}(r_0)$ is shown in Fig.~\ref{fig:SS2}.
}
\label{fig:SS1}
\end{figure*}

\begin{figure}[th]
\centering
\includegraphics[width=0.45\textwidth]{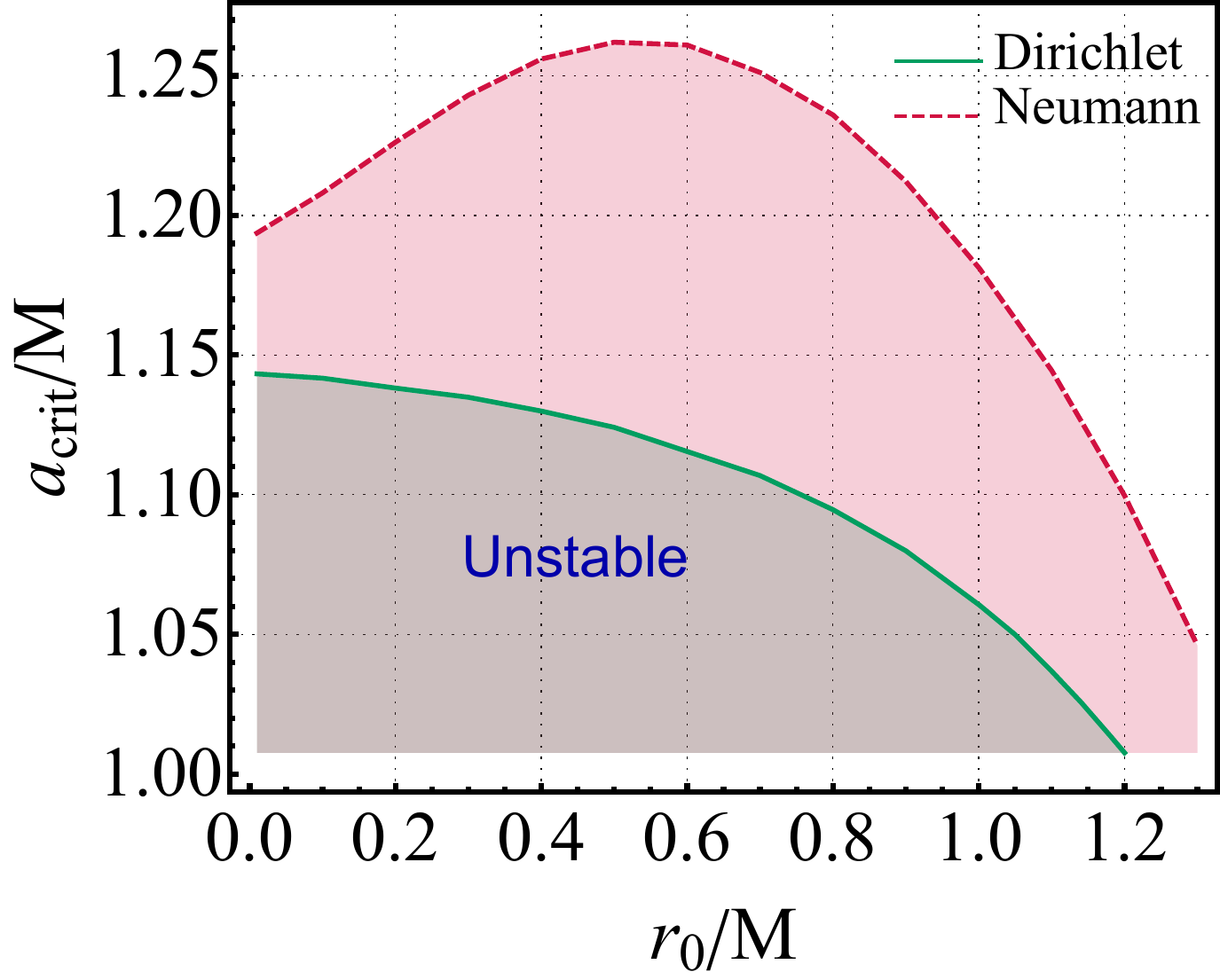}
\caption{Critical value of the spin below which a superspinning ($a>M$) ECO is unstable. We show the cases of Dirichlet and Neumann boundary conditions imposed at the surface $r_0$. Note that, when $a_{\rm crit}\to M$, the curves connect continuously to the corresponding critical value for $a<M$ (cf.\ Fig.~\ref{fig:crit}).
}
\label{fig:SS2}
\end{figure}

The fundamental mode for a model of superspinning ECO with Dirichlet boundary conditions imposed at $r=r_0$ is shown in Fig.~\ref{fig:SS1} for different choices of $r_0$.
(We checked that the Neumann case is qualitatively similar.)
As already found in Ref.~\cite{Pani:2010jz}, the mode is unstable only below a critical value of the spin, $a_{\rm crit}(r_0)$, which is shown in Fig.~\ref{fig:SS2}. This result, together with Fig.~\ref{fig:crit}, suggests that the ergoregion instability is \emph{not} effective for ECOs spinning either \emph{below} [$a/M<{\cal O}(0.1)$] or \emph{above} [$a/M>{\cal O}(1)$] some critical spin, whose exact value depends on the compactness of the object (cf.\ Figs.~\ref{fig:crit} and \ref{fig:SS2}).

%

%
\bibliographystyle{utphys}
\bibliography{refs}
\end{document}